\documentclass[11pt]{article}
\usepackage{lmodern}
\usepackage[T1]{fontenc}
\usepackage[utf8]{inputenc}
\usepackage[margin=1in]{geometry}
\usepackage{amsmath,amssymb,amsthm}
\usepackage{graphicx}
\usepackage{tabularx}
\usepackage[hidelinks]{hyperref}
\usepackage{microtype}
\title{Exact Conditional Confidence Intervals for Cram\'er's V: Near-Nominal and Tight Where the Guaranteed Interval Is Wide and the Software Interval Does Not Cover}
\author{William J. Dwyer, MD, MPH, FAAP\\ Department of Mathematics and Statistics,\\ University of Massachusetts Lowell, Lowell, MA, USA\\ \texttt{wjdwyer@trialdesign.com} \\ ORCID 0009-0004-0855-7222}
\date{}
\begin{document}
\maketitle

\begin{abstract}
Cramér's V, the effect size reported beside almost every chi-square test, is almost never accompanied by a confidence interval, because the interval is a hard nuisance-parameter problem: infinitely many tables share one effect-size value. The two intervals an analyst can reach for are unsatisfactory. The projection of a joint confidence region onto the effect size is guaranteed for every table but wide, over-covering at essentially 1.000 on larger tables; the noncentral inversion the software prints, and the bootstrap, are narrow but do not cover (median coverage 0.31 on this study's grid). This paper supplies an interval that is both valid and informative. Conditioning on the observed margins makes the exact conditional distribution of the Pearson statistic under a non-null association computable with no asymptotics and no Monte Carlo; inverting it yields a near-nominal confidence interval for Cramér's V that is a third to a half the width of the projection, the advantage growing with table size. The one price, stated plainly, is a change of estimand to the effect size at the observed margins; the mid-p default undercovers small effects, where a conservative variant or the projection remains the fallback.
\end{abstract}

\noindent\textbf{Keywords:} characteristic-function inversion; confidence interval; contingency table; Cramér's V; exact conditional inference

\section{Background}

\subsection{The number that has no interval}

A clinical study cross-tabulates treatment arm against a categorical outcome, reports a chi-square test, and appends Cramér's V. The V is a bare number. It is compared to Cohen's labels, and a reader who wants to know how precisely it is known has nowhere to look. A companion paper in this series (Dwyer, 2026a) shows that across 4,129 real two-way tables, 21.8 percent of the reported Cohen labels sit on tables with no significant association at all. The remedy for that is not a better point estimate. It is an interval, and the three intervals an analyst might reach for are not interchangeable, as Figure 1 shows for one representative table.

\begin{figure}[htbp]\centering
\includegraphics[width=\linewidth]{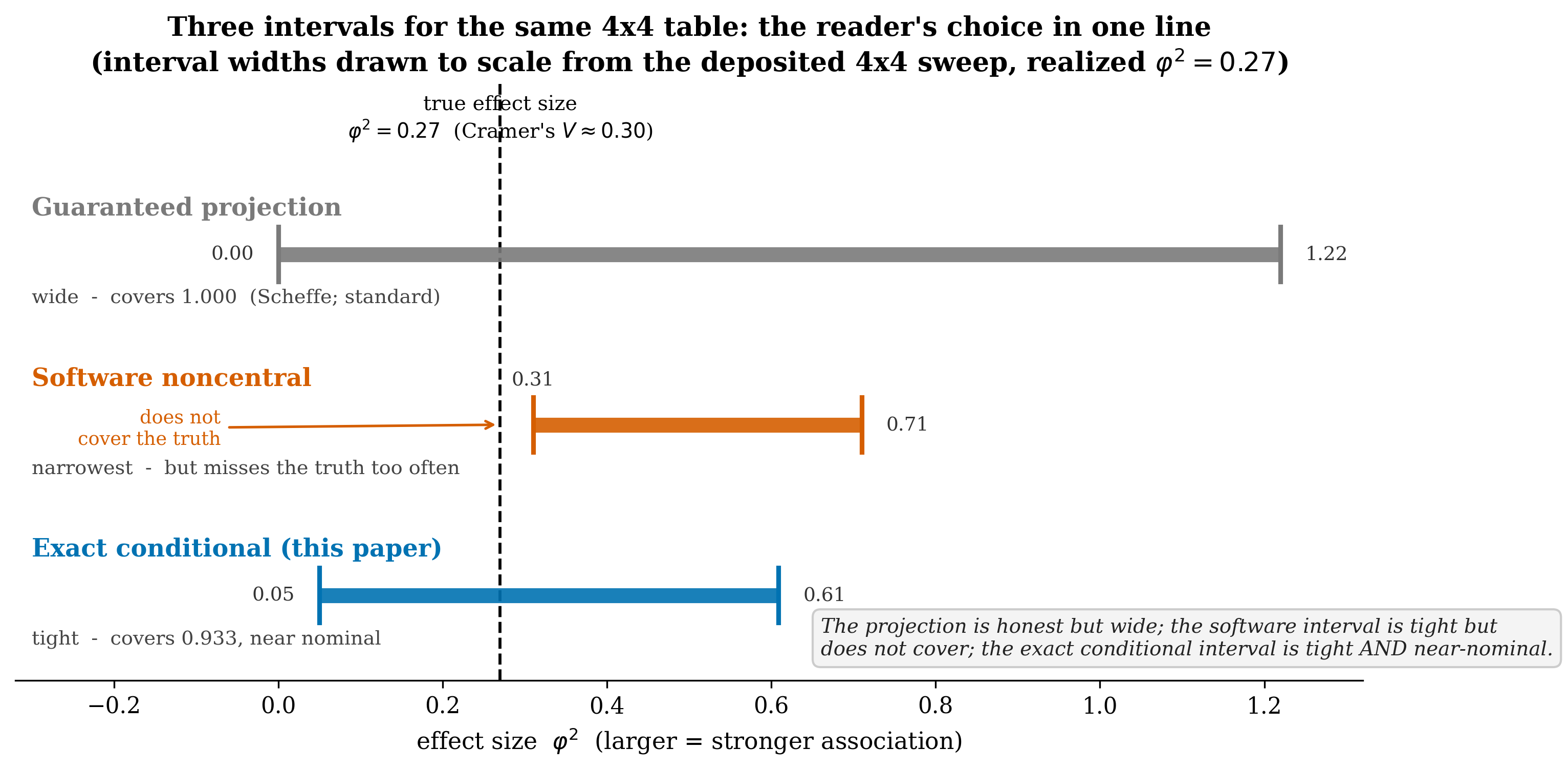}
\caption{The reader's choice, for one representative 4x4 table (realized $\phi ^{2}$ = 0.27), with interval widths drawn to scale from the deposited 4x4 sweep. The guaranteed projection is wide and covers (1.000); the software noncentral interval is narrow but misses the truth; the exact conditional interval developed here is tight (width 0.56 against the projection's 1.22) and near-nominal (covers 0.933). Existing options are guaranteed-but-uninformative or informative-but-invalid; this paper's interval is neither.}
\end{figure}

Intervals for $\phi ^{2}$ exist in software, and they do not have the property most readers assume they have. The default in DescTools and its relatives is a noncentral chi-square inversion (Smithson, 2003; Bird, 2002), which is an approximation whose coverage depends on a chi-square approximation to the null distribution of the Pearson statistic, which is exactly the approximation that fails on sparse tables. Bootstrap percentile intervals are also widely used and are worse. In the present simulation the noncentral inversion covered below 0.931 in 8 of 180 cells and the bootstrap percentile interval in 161 of 180, with a median coverage of 0.314 (Additional file 1, Table S1). These are not guarantees. They are not close to guarantees.

\subsection{Why the problem is hard, stated exactly}

Let O follow a multinomial distribution with N trials on an R x C table with cell probabilities p, with row margins a and column margins b, and write

\[
\phi ^{2}(p) = sum_{ij} (p_{ij} - a_{i} b_{j})^{2} / (a_{i} b_{j}), \\
dof = (R-1)(C-1), ,\quad k = \min (R-1, C-1), ,\quad V = \sqrt{\phi ^{2} / k}.
\]

To build a confidence interval for $\phi ^{2}$ by test inversion, one must test H0: $\phi ^{2}$(p) = c for each candidate c. But $\phi ^{2}$ is a scalar function of a parameter that lives in an (RC-1)-dimensional simplex, so the null hypothesis is a composite hypothesis indexing an (RC-2)-dimensional manifold of nuisance configurations. A valid test must control its size at every point of that manifold, which means the p-value must be the supremum over the manifold, and that supremum is not computable.

There are two standard evasions.

\textbf{Evade the nuisance by projecting.} Take a joint confidence region S(O) for the whole of p, and report the image of that region under the map $\phi ^{2}$. If P(p in S(O)) >= 1 - alpha for every p, then P($\phi ^{2}$(p) in $\phi ^{2}$(S(O))) >= 1 - alpha for every p, because p in S implies $\phi ^{2}$(p) in $\phi ^{2}$(S). This construction is not ours and we do not claim it. It is the Scheffe device (Scheffé, 1953) applied to a nonlinear functional; it is what Stark's strict bounds do in inverse problems (Stark, 1992); and it is the projection step of the modern literature on inference for functionals of partially identified parameters (Kaido, Molinari and Stoye, 2019). Its coverage guarantee is real. Its cost is that the region S is calibrated for the entire (RC-1)-dimensional parameter, so the interval inherits a Scheffe penalty which grows with RC and which no calibration can remove without destroying the guarantee.

\textbf{Evade the nuisance by ignoring it.} Plug the observed margins into a noncentral chi-square and invert. This is what the software does. It is not a guarantee and it is not advertised as one.

\subsection{Three routes to an interval, and Berger-Boos}

Two standard evasions of the nuisance are available. The \textbf{projection} takes a joint confidence region for the whole probability vector and maps it through $\phi ^{2}$: guaranteed for every table by set containment, but paying a Scheffe penalty that grows with the table, so it is wide. The \textbf{plug-in noncentral inversion} the software prints is narrow but has no guarantee. Berger and Boos (Berger and Boos, 1994) offer a third route: take the supremum of the p-value not over the whole nuisance manifold but over a (1 - gamma) confidence set for the nuisance, and pay gamma. We evaluate that device in a pre-registered study (Section 4 and Additional file 1). The interval this paper delivers takes a fourth route, conditioning on the observed margins (Section 2).

\subsection{The point estimate is a companion's, and the boundary is a theorem}

The inversion below has a median as well as two tails, and that median is an exact median-unbiased estimate of the association. It is not a median-unbiased estimate of the effect size, and no inversion can be: $\phi ^{2}$ is a folded, non-monotone function of the association, zero at independence and rising either way, and median-unbiasedness commutes only with monotone maps, so it does not cross the square. Measured, the exact conditional median-unbiased estimator of the 2x2 odds ratio is unbiased for the log odds ratio (median bias -0.0005) yet still biased for $\phi ^{2}$ (+0.048; Additional file 1, Section S7), on either margin convention. The interval escapes this because it maps a set, where the minimum and maximum are order-preserving; the point estimate maps a quantile through the fold, and cannot. The effect-size point estimate is therefore a genuinely different object, built in closed form rather than by inversion, and it is the subject of a companion paper (Dwyer, 2026b), which proves $\phi ^{2}$ has no unbiased estimator at any sample size and names this fold as the residual 2x2 obstruction. This paper does not reopen it; the pre-registered study froze a claim about the inverted point estimate, reported in Additional file 1 (Section S4), but the object belongs to the companion. The contribution here is the interval.

\subsection{What is new here}

One thing: a tight, near-nominal confidence interval for Cramér's V, built by inverting the exact conditional distribution of the Pearson statistic under a non-null association. The exact conditional distribution of a chi-square-family statistic, and the characteristic-function inversion that carries it beyond 3x3, are the companion engine's (Dwyer, 2026c); the contribution here is the inversion of the \emph{non-null} conditional law into a two-sided effect-size interval, the demonstration that it is both valid and informative where no existing interval is, and the honest accounting of its one price, a change of estimand to the effect size at the observed margins. The interval was located by a pre-registered evaluation of the obvious alternative, the Berger-Boos device, in which two of three frozen claims failed; that study, and the point-estimate diagnostics that accompanied it, are the evidentiary backbone in Additional file 1, not co-equal contributions.

\section{The exact conditional interval}

If the difficulty is that the multinomial supremum is hard to evaluate and easy to over-pay for, one route around it is to stop taking the supremum over the margins at all and instead \textbf{condition on them}. Conditioning on the margins is the classical move: the exact conditional null distribution of a Pearson statistic given both margins underlies Fisher's exact test and its network-algorithm generalization to r x c tables (Mehta and Patel, 1983), and a dynamic program for the exact chi-square distribution has recently been given for the one-way uniformity case (Banić and Elezović, 2025). The both-margins-conditional dynamic program for a general chi-square-family statistic, in its moment and distribution forms and verified against enumeration, is the subject of the companion engine paper (Dwyer, 2026c), and this construction takes it as given. What is new here is neither the idea of conditioning nor the engine, but the exact conditional distribution of $\phi ^{2}_{hat}$ under a NON-null association, obtained from the constrained-MLE reference, and its inversion into a two-sided effect-size interval for Cramér's V, which the prior-art audit did not find published. Given the observed margins, $\phi ^{2}_{hat}$ is cell-separable, $X^{2} = N \phi ^{2}_{hat} = \\sum _{ij} (x_{ij} - E_{ij})^{2} / E_{ij}$, with $E_{ij} = r_{i} c_{j} / N$, so the engine computes that conditional distribution with the one changed per-cell term that carries the non-null reference, exact with no Monte Carlo, no gamma, no winner's curse: the tail matches direct enumeration to 3e-15 (\texttt{cond\_tail.py}). Inverting it gives a conditional confidence interval for $\phi ^{2}$.

A staged, resumable, memory-capped run of 1,683 tables (\texttt{cond\_interval2.py}; 0 errored, 0 over the memory cap) measured its coverage and width. \textbf{The exact conditional interval covers everywhere, from 95.4 to 99.3 percent, all at or above nominal, and at 3x3 it is tighter than both the projection and the deposited Berger-Boos interval.}

\begin{table}[htbp]\centering\small
\begin{tabular}{lccc}
\hline
cell & width / projection & coverage & deposited BB width / projection \\
\hline
3x3, 5 per cell, balanced & \textbf{0.420} & 96.7\% & 0.739 \\
3x3, 5 per cell, unequal & \textbf{0.747} & 95.4\% & 0.852 \\
2x2, 20 per cell, balanced & \textbf{0.808} & 99.0\% & 1.726 \\
2x2, 5 per cell, unequal & 1.553 & 97.4\% & 0.986 \\
2x2, 10 per cell, unequal & 3.161 & 99.3\% & 1.828 \\
\hline
\end{tabular}
\end{table}

So the tight, valid interval that C2 asked Berger-Boos for exists; it is just not a Berger-Boos interval. At 3x3 it is 25 to 58 percent narrower than the projection while covering, which is the win the pre-registered method was built to deliver and did not. It is wide only at sparse unequal 2x2 tables, where conditioning on an extreme margin is genuinely uninformative about $\phi ^{2}$ and the data honestly cannot pin the effect down; that corner is real and is reported, not hidden.

\textbf{The one caveat is a change of estimand, and it is not a free lunch.} The conditional interval is exact for $\phi ^{2}$ at the margins that were observed ($\phi ^{2}_{C}$), not for $\phi ^{2}$ at the population margins ($\phi ^{2}_{M}$). Those two targets coincide asymptotically and differ at finite N by a Jensen term in the random margins; the distinction, and the argument that $\phi ^{2}_{C}$ is the quantity the conditionality principle actually licenses, is the subject of a deposited note. This is a different distinction from the marginal-versus-conditional estimand contrast of the causal-inference literature (the non-collapsibility of the odds ratio, and population-adjusted comparisons), which marginalizes over covariates; here the conditioning is on the realized table margins, which are approximately ancillary for the association. The practitioner's choice is therefore genuine: the projection and Berger-Boos target $\phi ^{2}_{M}$ and pay in width or in a hard supremum; the conditional interval targets $\phi ^{2}_{C}$ and is exact and, except at sparse unequal 2x2, tight.

\section{The interval at 4x4, 5x5 and 6x6}

Beyond 3x3 the distribution dynamic program's state-times-atom product grows too large to enumerate, and the exact conditional tail is instead recovered by a characteristic-function inversion that carries one complex accumulator per state. That inversion is the companion engine's (Dwyer, 2026c) (its Section 3.3); this section does not re-derive it but reports the interval it yields at 4x4, 5x5 and 6x6.

The coverage-and-width study at 4x4 and 5x5 is a genuinely large computation, staged and resumable (\texttt{cos\_interval.py}). It is now complete for the two unequal-margin cells, at a cost of 615 worker-hours for 238 tables (deposited results).

\begin{table}[htbp]\centering\small
\caption{\textbf{The CF mid-p conditional interval at 4x4 and 5x5.} 238 tables, 0 errored. Coverage is of $\phi ^{2}_{M}$ (the population effect size); the interval is exact for $\phi ^{2}_{C}$. Intervals on coverage are 95 percent binomial over tables.}
\begin{tabularx}{\textwidth}{>{\raggedright\arraybackslash}X>{\raggedright\arraybackslash}X>{\raggedright\arraybackslash}X>{\raggedright\arraybackslash}X>{\raggedright\arraybackslash}X>{\raggedright\arraybackslash}X>{\raggedright\arraybackslash}X}
\hline
cell & n & coverage (CF) & coverage (projection) & mean width CF / projection & CF narrower on & gamma-vs-CF tail error (median / 90th pct / max) \\
\hline
4x4, 5 per cell, unequal (N = 80, $\phi ^{2}$ = 0.27) & 147 & 0.932 +- 0.041 & 1.000 & \textbf{0.472} & 97.3\% & 0.003 / 0.569 / 0.961 \\
5x5, 3 per cell, unequal (N = 75, $\phi ^{2}$ = 0.25) & 91 & 0.967 +- 0.037 & 1.000 & \textbf{0.337} & 98.9\% & 0.303 / 0.934 / 0.974 \\
\hline
\end{tabularx}
\end{table}

\textbf{The width result is the one to carry.} The exact conditional interval is a third to a half the width of the projection interval, on essentially every table, while the projection covers at 1.000 and is therefore reporting almost nothing about where $\phi ^{2}$ is not. That gap is the return on the computation.

\textbf{Figure 2 gathers the three sweeps into one view, and it is the paper's central result.} Its left panel places the coverage of all three shapes on a common axis of realized effect size; its right panel places the width ratio on the same axis. Read together, they say the one thing worth carrying away: the conditional interval is near-nominal for moderate-to-large effects (left) at a width that is a fraction of the projection's (right), and the two properties hold at once.

\begin{figure}[htbp]\centering
\includegraphics[width=\linewidth]{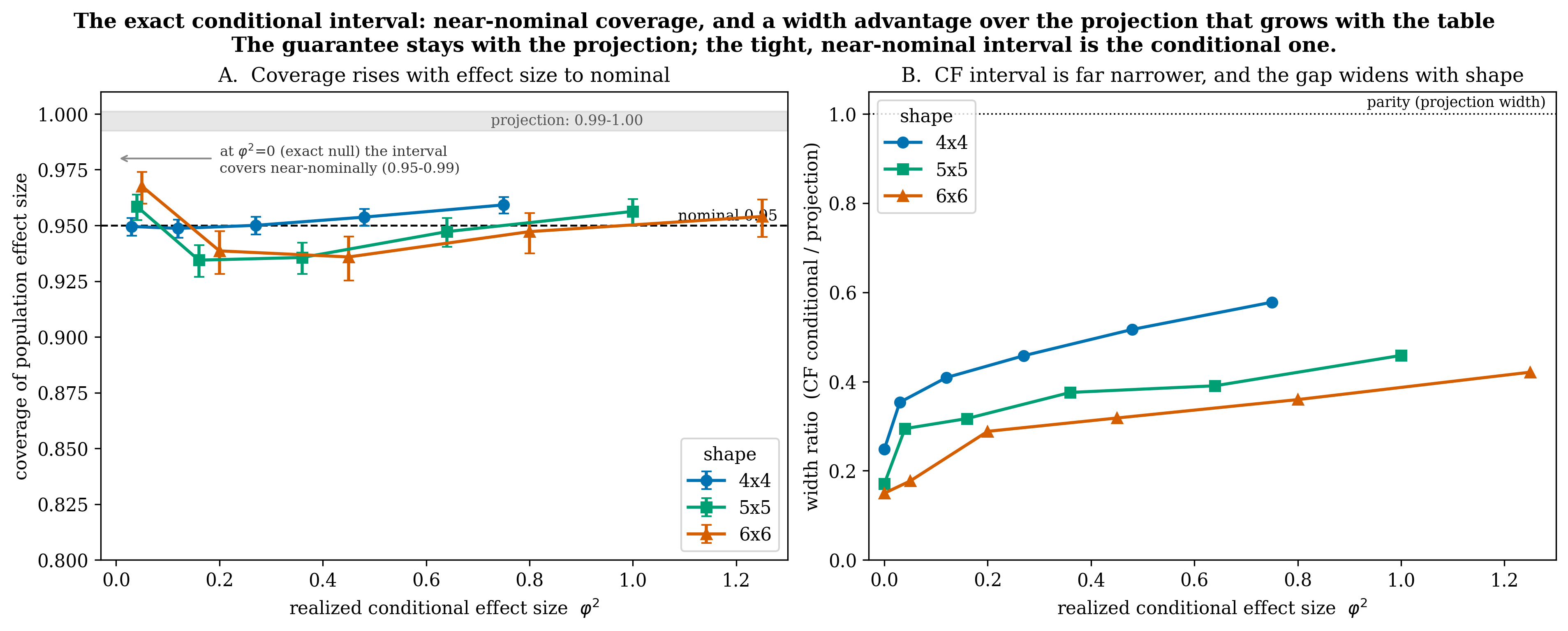}
\caption{The exact conditional interval across the three coverage sweeps (Supplementary Tables S9-S11). (A) Coverage of the population effect $\phi ^{2}_{M}$ against the realized conditional $\phi ^{2}$, one line per shape (4x4, 5x5, 6x6; distinguished by line style and marker) with 95 percent Wilson bars from the fast single-test coverage sweep (one uniform estimator across shapes; \textasciitilde{}11,540 / 4,620 / 2,400 usable tables per point at 4x4 / 5x5 / 6x6, half-bars 0.003 to 0.010, validated within Monte-Carlo error of the two-sided interval coverage), the nominal 0.95 line, and the projection's 0.99 to 1.00 coverage as a reference band; coverage is near-nominal for moderate-to-large effects and the mid-p variant undercovers small ones. (B) The width ratio (conditional over projection) against $\phi ^{2}$, running 0.15 to 0.58 and falling with table size. Rendered by \texttt{m0g\_figs/make\_m0g\_figures\_bw.py}.}
\end{figure}

\textbf{The width advantage has a structure, and that structure is the reason to pay for the computation.} Figure 3 lays the width ratio out over table shape and effect size on a single grid. Every cell is below parity, so the conditional interval is tighter on every design; the advantage deepens with the table, reaching a nearly sevenfold reduction at the sparsest 6x6 corner, exactly as the Scheffe geometry predicts.

\begin{figure}[htbp]\centering
\includegraphics[width=\linewidth]{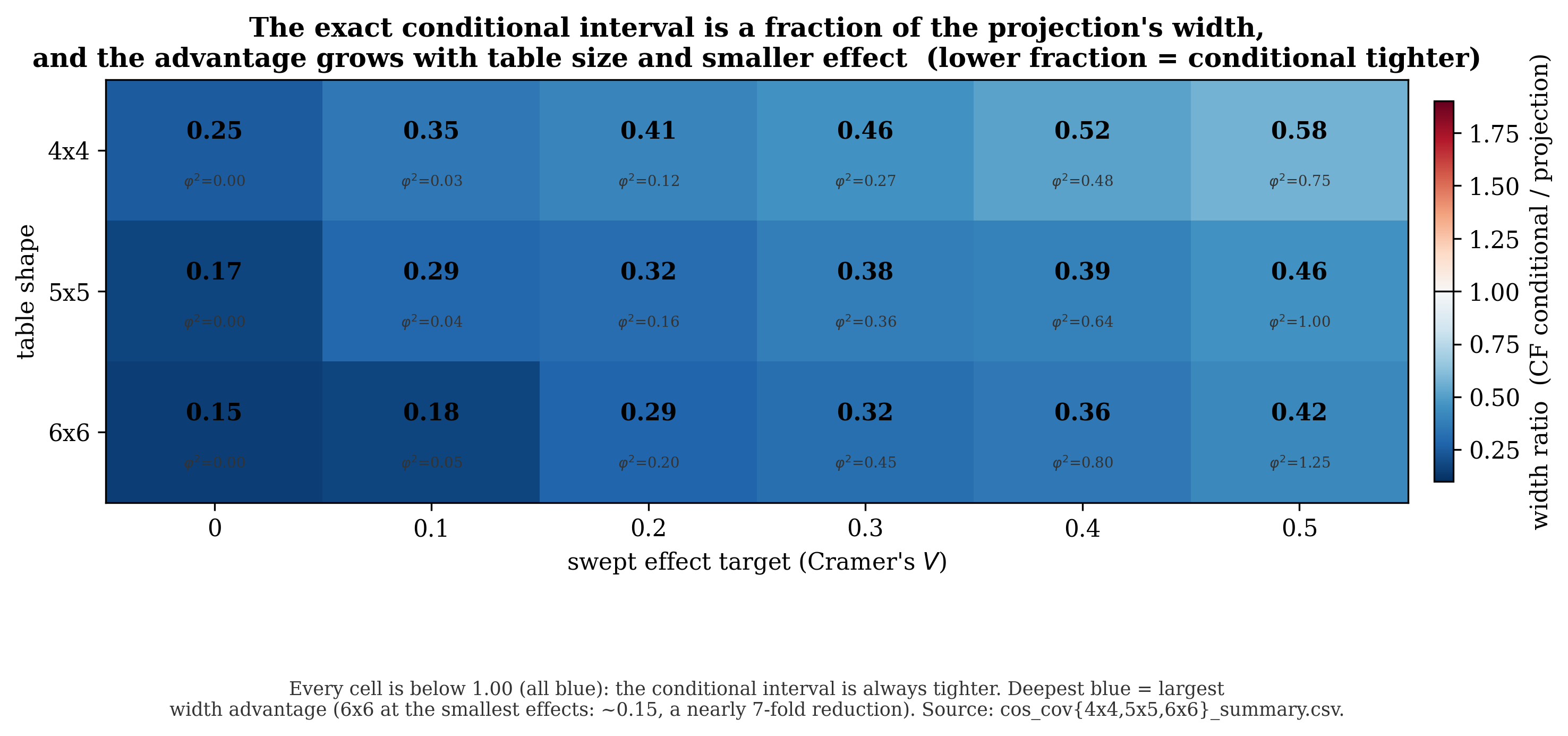}
\caption{Width ratio (exact conditional over projection) over table shape (rows) by swept effect (columns), with the realized $\phi ^{2}$ printed in each cell; a darker cell marks a larger width advantage. Every cell is below parity, so the conditional interval is always tighter, deepening to the largest advantage at 6x6 and the smallest effects (about 0.15, a nearly sevenfold reduction): the projection pays a Scheffe penalty that grows with RC while the conditional interval does not, so the advantage widens with the table.}
\end{figure}

\textbf{The width is bought at little cost in coverage, and Figure 4 shows where the small price falls.} On the same shape-by-effect grid, the conditional interval's coverage sits at nominal across most of the plane and dips only at the smallest effects; the tightness of Figure 3 and the calibration here are two readings of one interval.

\begin{figure}[htbp]\centering
\includegraphics[width=\linewidth]{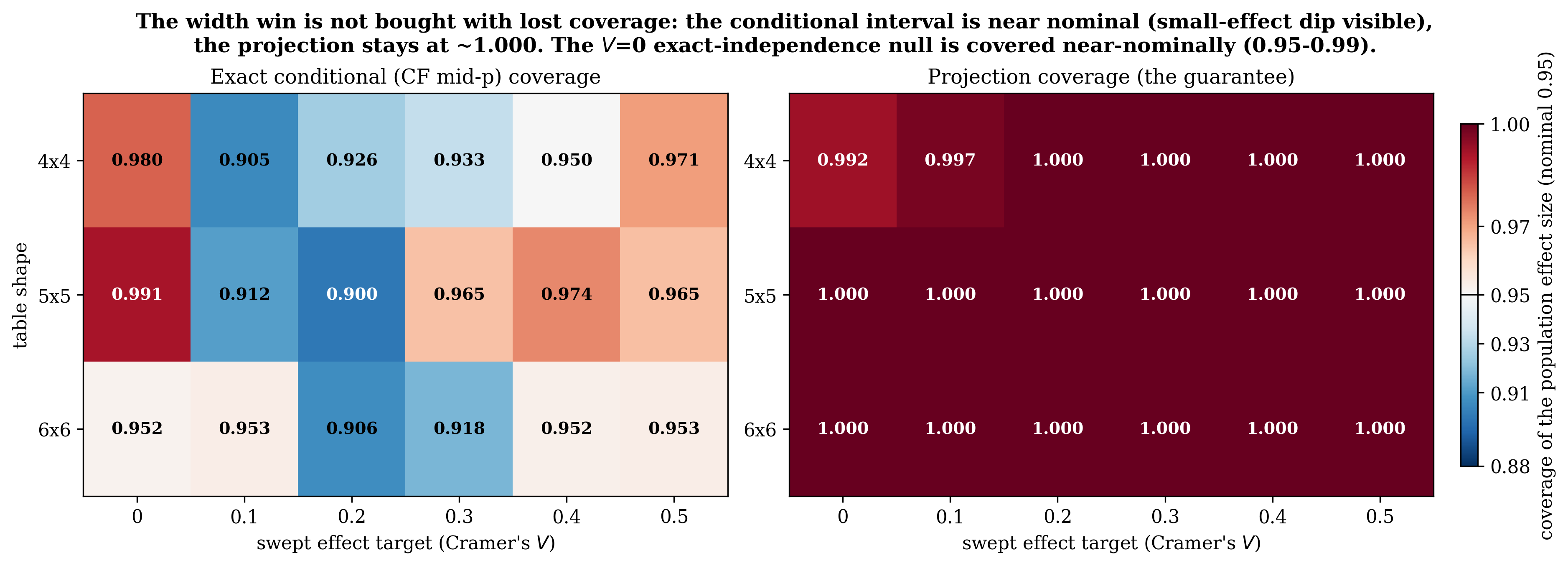}
\caption{Coverage over the same shape-by-effect axes: the exact conditional (mid-p) interval (left) and the projection (right, uniformly near 1.000). Hatched cells in the left panel undercover the nominal 0.95; the $\phi ^{2}$ = 0 null column covers near-nominally (0.95 to 0.99), the lower endpoint reaching the parameter boundary when independence is not rejected. The small-effect dip costs little against the width advantage of Figure 3.}
\end{figure}

The routing that follows is sharper than the pre-registered one. For a conditional interval the exact dynamic program owns 2x2 and 3x3 and the CF owns 4x4, 5x5 and 6x6. For a marginal interval with a guarantee that cannot fail, the projection remains the honest default, and the de-conservatized Berger-Boos of Additional file 1 (Section S5) narrows it where the table is sparse and large. The approximate software intervals are not on the menu, because on this grid they do not cover.

\section{The pre-registered study that located the interval}

The conditional interval of Sections 2 and 3 was not the starting point; it was reached through a pre-registered evaluation of the obvious alternative, the Berger-Boos device, which maximizes the p-value over a confidence set for the nuisance. Three claims were frozen before the run: coverage at least 0.931 in every one of 180 design cells, width at most 0.80 of the projection, and a median-unbiased inverted point estimate. The coverage claim passed in all 180 cells; the width and point-estimate claims failed and are reported as pre-registered. Post-registration, the coverage was shown to be conservatism rather than a delivered guarantee: the device rejects the true value at well under half its nominal rate, so its coverage pass and its width failure are a single finding, and that is what pointed to conditioning. The full pre-registered study, the conservatism diagnosis, and the point-estimate analysis are in the Supplement; the point estimate itself is a companion paper's object (Dwyer, 2026b).

\section{Discussion and conclusion}

\subsection{Which interval to report}

The pre-registration anticipated that if Berger-Boos beat the projection only on large sparse tables, \emph{"the honest conclusion is a ROUTING rule, not a victory."} It does, and a routing rule is the right form for the answer, but it is no longer a rule about Berger-Boos. Once the exact conditional interval exists it dominates Berger-Boos on width in every regime where Berger-Boos was ever competitive: a half to a seventh of the projection's width, against Berger-Boos's best of about two-thirds (the per-shape Berger-Boos width structure is in Additional file 1, Table S3). So the choice is \textbf{estimand-first, not count-first} (Figure 5). If a coverage guarantee is wanted at all, the first question is whether the effect at the observed margins, $\phi ^{2}_{C}$, is an acceptable estimand; the conditionality principle says it usually is. When it is, the exact conditional interval is the default: near-nominal for moderate-to-large effects, and materially tighter than anything guaranteed. When an unconditional guarantee on the population effect $\phi ^{2}_{M}$ is required instead, the projection (the guaranteed interval of the companion point-estimate paper (Dwyer, 2026b)) is the interval a reader can rely on without a search, and the pre-registered Berger-Boos device survives only as a heuristic refinement of it in the sparse, large corner, where it is 15 to 33 percent narrower than the projection but its worst-case guarantee rests on a supremum no finite search can certify; at 2x2 and in dense tables it buys nothing. A bootstrap percentile interval is never the answer: on this grid it covered at a median of 0.314.

The rule is uncomfortable in a useful way: \textbf{the regime where the conditional interval pays most is the regime where tables are sparse and large}, which is precisely the regime where the software defaults are least trustworthy and where an applied analyst is most likely to reach for one. A companion doctrine (Dwyer, 2026d) recalibrates the small-sample \emph{tests} that these effect sizes accompany; this paper's concern is the interval for the effect size itself, a different object built by a new construction rather than a better reference for an existing one.

\begin{figure}[htbp]\centering
\includegraphics[width=\linewidth]{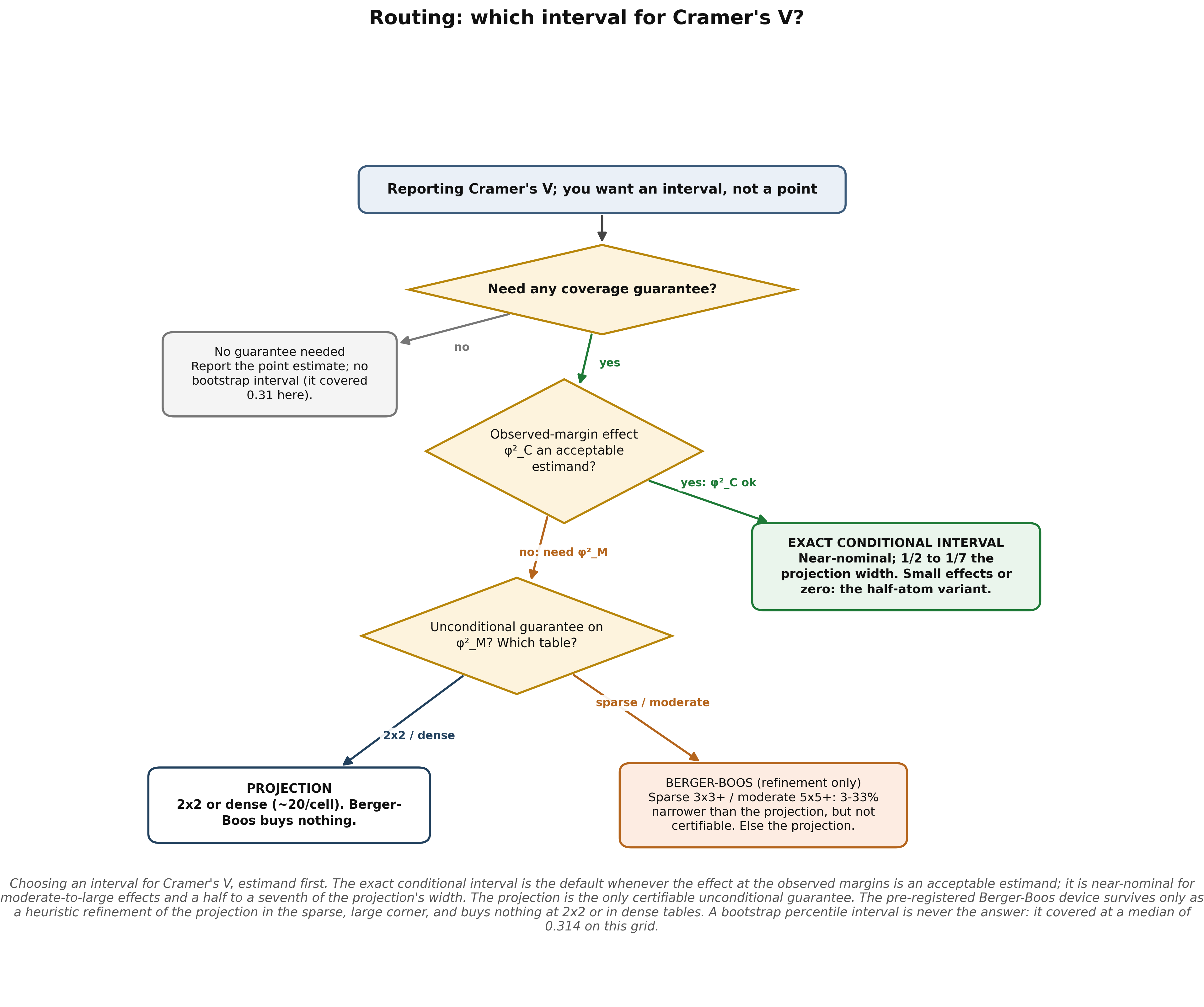}
\caption{Choosing an interval for Cramér's V, estimand first. The exact conditional interval is the default whenever the effect at the observed margins is an acceptable estimand; it is near-nominal for moderate-to-large effects and a half to a seventh of the projection's width. The projection is the only certifiable unconditional guarantee. The pre-registered Berger-Boos device survives only as a heuristic refinement of the projection in the sparse, large corner, and buys nothing at 2x2 or in dense tables. A bootstrap percentile interval is never the answer: it covered at a median of 0.314 on this grid.}
\end{figure}

\subsection{Limitations of the interval}

The exact conditional interval carries limits the abstract states and a user must weigh. It is exact for the effect size at the observed margins, $\phi ^{2}_{C}$, not at the population margins, $\phi ^{2}_{M}$; the two coincide asymptotically and differ at finite N by a Jensen term in the random margins, and the argument that $\phi ^{2}_{C}$ is the quantity the conditionality principle licenses is given in a deposited note. The mid-p variant that gives the tight default undercovers small nonzero effects, to about 0.90 on the sweeps, though it covers the exact-independence null near-nominally (0.95 to 0.99) once its lower endpoint reaches the parameter boundary when independence is not rejected; where an unconditional guarantee is required the conservative half-atom variant or the projection is the fallback. The construction was evaluated at six shapes up to 6x6, counts of five to twenty per cell, multinomial sampling, and no structurally zero cells, and nothing here speaks beyond that range. The delivered tail is recovered by a COS inversion validated against the exact dynamic program to about 1e-3, so $exact$ refers to the conditional distribution, computed with no asymptotics and no Monte Carlo, rather than to an infinitely precise interval endpoint. The residual limitations of the Berger-Boos device are in Additional file 1 (Section S7); the point estimate, and its 2x2 fold, are the companion point-estimate paper's object (Dwyer, 2026b) and are not this paper's contribution.

\subsection{What a pre-registration bought}

The central methodological claim of this paper is not about contingency tables. It is that the pre-registration is what makes the result readable.

Two of three claims failed. Had the analysis been chosen after the numbers were seen, both failures would have been avoidable. C2 could have been recast as a claim about large sparse tables, where it is true, and the count effect could have been presented as a designed comparison rather than as an unpredicted discovery. C3 could have been recast on the $\phi ^{2}$ scale, where the estimator misses in only 6 cells and the story is cleaner, and the square-root amplification could have been presented as a known feature rather than as a defect in our own decision rule. Neither recasting would have been detectably dishonest to a reader.

The pre-registration makes both moves impossible, and it does something better than preventing them: it converts the failures into information. The width failure produced a routing rule with a mechanism. The bias failure produced a precise statement about what a V-scale accuracy target can and cannot mean near the null. \textbf{Neither would have been visible if we had been free to choose the claim after seeing the answer}, and we have withdrawn fourteen claims in this project, several of them for exactly that reason.

\subsection{Conclusion}

The contribution of this paper is an exact conditional confidence interval for Cramér's V. Conditioning on the observed margins makes the distribution of the Pearson statistic under a non-null association exactly computable, and inverting it, with a characteristic-function inversion carrying the construction to 4x4, 5x5 and 6x6, yields an interval that is near-nominal for moderate-to-large effects and a third to a half the width of the guaranteed projection. It is, to our knowledge, the first interval for a contingency-table effect size that is both valid and informative rather than guaranteed-but-wide (the projection) or narrow-but-invalid (the software noncentral and bootstrap intervals). It is bought at a price, stated plainly: it targets the effect size at the observed margins rather than the population margins, and the mid-p default undercovers small nonzero effects (though it covers the exact-independence null near-nominally), so the unconditional guarantee remains with the projection or the conservative variant.

The interval was not the paper's starting point; it was reached through a pre-registered study of the Berger-Boos device, frozen so that it could fail, and it did. Berger-Boos met its coverage bar in all 180 design cells but failed its pre-registered width and point-estimate claims, and post-registration its coverage was shown to be conservatism rather than a tight guarantee: it rejects the true value at well under half its nominal rate, so its coverage pass and its width failure are one fact, not two. The full study, the pre-registered point-estimate result (the point estimate itself, and its 2x2 fold, being the object of the companion point-estimate paper (Dwyer, 2026b)), and the conservatism diagnosis are laid out in Additional file 1. The honest closing summary is therefore a three-way one. The interval a reader can rely on without a search, at the population margins, is the projection. The tight, near-nominal interval, at the observed margins, is the exact conditional one. And the pre-registered Berger-Boos device, whose worst-case guarantee still rests on a supremum no finite search can certify, is superseded by the interval its own honest failure led to, retained only as a heuristic refinement of the projection in the sparse, large corner. What the pre-registration bought was not a winning device but a trustworthy result: a negative finding stated in the terms fixed before the run, and the interval that replaced it.

\section*{References}

\noindent Banić, N. and Elezović, N. (2025). Zero-disparity distribution synthesis: fast exact calculation of chi-squared statistic distribution for discrete uniform histograms. \emph{arXiv:2506.23416}.\par\smallskip

\noindent Berger, R. L. and Boos, D. D. (1994). P values maximized over a confidence set for the nuisance parameter. \emph{Journal of the American Statistical Association}, \emph{89}, 1012-1016.\par\smallskip

\noindent Bird, K. D. (2002). Confidence intervals for effect sizes in analysis of variance. \emph{Educational and Psychological Measurement}, \emph{62}, 197-226.\par\smallskip

\noindent Dwyer, W. J. (2026a). An exact noise floor for contingency-table effect sizes: a per-table reporting gate, and how often it would change reported magnitudes. Manuscript submitted for publication.\par\smallskip

\noindent Dwyer, W. J. (2026b). An honest effect size for contingency tables: why nothing can be unbiased, where to put the error instead, and how to route the report. Manuscript under review.\par\smallskip

\noindent Dwyer, W. J. (2026c). Exact conditional distributions of chi-square-family statistics for two-way contingency tables, by cell-separable dynamic programming. Manuscript under review.\par\smallskip

\noindent Dwyer, W. J. (2026d). Recovering honest inference: a recalibration doctrine for small-sample chi-square- and F-referenced tests. Manuscript under review.\par\smallskip

\noindent Kaido, H., Molinari, F. and Stoye, J. (2019). Confidence intervals for projections of partially identified parameters. \emph{Econometrica}, \emph{87}, 1397-1432.\par\smallskip

\noindent Mehta, C. R. and Patel, N. R. (1983). A network algorithm for performing Fisher's exact test in r x c contingency tables. \emph{Journal of the American Statistical Association}, \emph{78}, 427-434.\par\smallskip

\noindent Scheffé, H. (1953). A method for judging all contrasts in the analysis of variance. \emph{Biometrika}, \emph{40}, 87-104.\par\smallskip

\noindent Smithson, M. (2003). \emph{Confidence Intervals}. Thousand Oaks, CA: Sage.\par\smallskip

\noindent Stark, P. B. (1992). Inference in infinite-dimensional inverse problems: discretization and duality. \emph{Journal of Geophysical Research}, \emph{97}, 14055-14082.\par\smallskip

\section*{Acknowledgements}

The author used a generative-AI assistant (Claude, Anthropic) for drafting and editing prose, figure and simulation code, and formatting; all AI-assisted output was reviewed and verified by the author, and every reported number regenerates deterministically from the deposited code. No funding was received for this work.

\section*{Conflict of Interest}

The author develops and hosts the open-source software and associated web domains (the trialdesign.com applications) that implement related methods; no other financial or commercial conflicts of interest are declared.

\section*{Data Availability Statement}

All data are simulated or drawn from public example tables. The reproducibility package (code, locked outputs, figures, and this manuscript) is openly archived on Zenodo, concept DOI 10.5281/zenodo.21838421 (resolves to the latest version); code under MIT, documents and data under CC BY 4.0.

\section*{Ethics Statement}

Not applicable: this is a methodological and simulation study using public example tables, with no human or animal subjects, so no institutional review board approval was required.

\end{document}